\documentclass[review,3p]{elsarticle}

\biboptions{sort&compress}

\usepackage{numcompress}
\usepackage{color}
\usepackage[bookmarksnumbered,
    colorlinks,allcolors=blue,
]{hyperref}

\usepackage{amsmath}
\usepackage{amssymb}

\let\bm\boldsymbol% bold vector
\newcommand\rmd{\mathrm{d}}% derivative
\newcommand\rme{\mathrm{e}}% exponential / electron
\newcommand\rmi{\mathrm{i}}% imaginary unit / ion
\newcommand\rms{\mathrm{s}}% sound (speed)
\newcommand\pd{\partial}% partial derivative
\newcommand\pdif[3][\empty]{\ifx#1\empty\frac{\pd#2}{\pd#3}\else\frac{\pd^#1#2}{\pd#3^#1}\fi}
\newcommand\lap{\Delta}% Laplacian
\newcommand\calL{\mathcal{L}}% linear
\newcommand\calN{\mathcal{N}}% nonlinear
\journal{Communications in Nonlinear Science and Numerical Simulation}

\begin{document}

\begin{frontmatter}



\title{Nonlinear ion acoustic waves scattered by vortexes}%


\author{Yuji Ohno\corref{cor}}%
\ead{ohno@ppl.k.u-tokyo.ac.jp}%
\cortext[cor]{Corresponding author}%

\author{Zensho Yoshida}%

\address{Graduate School of Frontier Sciences, The University of Tokyo, 5-1-5 Kashiwanoha, Kashiwa, Chiba 277-8561, Japan}%

\begin{abstract}
The Kadomtsev--Petviashvili (KP) hierarchy is the archetype of infinite-dimensional integrable systems,
which describes nonlinear ion acoustic waves in two-dimensional space.
This remarkably ordered system resides on a singular submanifold (leaf) embedded in a larger phase space
of more general ion acoustic waves (low-frequency electrostatic perturbations).
The KP hierarchy is characterized not only by small amplitudes but also by irrotational (zero-vorticity) velocity fields.
In fact, the KP equation is derived by eliminating vorticity at every order of the reductive perturbation.
Here we modify the scaling of the velocity field so as to introduce a vortex term.
The newly derived system of equations consists of a generalized three-dimensional KP equation and a two-dimensional vortex equation.
The former describes `scattering' of vortex-free waves by ambient vortexes that are determined by the latter.
We say that the vortexes are `ambient' because they do not receive reciprocal reactions from the waves
(i.e., the vortex equation is independent of the wave fields).
This model describes a minimal departure from the integrable KP system.
By the Painlev\'e test, we delineate how the vorticity term violates integrability, bringing about an essential three-dimensionality to the solutions.
By numerical simulation, we show how the solitons are scattered by vortexes and become chaotic.

%Highlight
%\begin{itemize}
%\item Generalizing the KP equation, a new system is formulated to include a vorticity.
%\item The integrability of the KP equation is broken by the introduction of a vortex term.
%\item Scattered by the vortex, the KP solitons undergo three-dimensional deformation.
%\item Beyond a threshold intensity of the vorticity, the scattered waves become chaotic.
%\item The Painlev\'e test delineates how the KP hierarchy is embedded in the new system.
%\end{itemize}

\end{abstract}


\end{frontmatter}

%%%%%%%%%%%%%%%%%%%%%%%%%%%%%%%%%%%%%%%%%%%%%%%%%%%%%%%%%%%%%%%%
%%%%%%%%%%%%%%%%%%%%%%%%%%%%%%%%%%%%%%%%%%%%%%%%%%%%%%%%%%%%%%%%
%%%%%%%%%%%%%%%%%%%%%%%%%%%%%%%%%%%%%%%%%%%%%%%%%%%%%%%%%%%%%%%%
\section{Introduction}
%%%%%%%%%%%%%%%%%%%%%%%%%%%%%%%%%%%%%%%%%%%%%%%%%%%%%%%%%%%%%%%%
%%%%%%%%%%%%%%%%%%%%%%%%%%%%%%%%%%%%%%%%%%%%%%%%%%%%%%%%%%%%%%%%
%%%%%%%%%%%%%%%%%%%%%%%%%%%%%%%%%%%%%%%%%%%%%%%%%%%%%%%%%%%%%%%%

%% IAW
The ion acoustic waves (IAWs) serve as a rich source of nonlinear phenomena.
The combination of the nonlinearity (by fluid convection) and the dispersion (by the nonlocal electric interactions)
enables IAWs to produce various structures ranging from order (such as solitons) to chaos (turbulences).
At the simplest one-dimensional geometry, small-amplitude IAWs become solitons;
\citet{Washimi1966:KdV} derived the Korteweg--de Vries (KdV) equation by the reductive perturbation method.
The Kadomtsev--Petviashvili (KP) equation \cite{Kadomtsev-Petviashvili} is a two-dimensional generalization of the KdV equation,
which was picked up by the `Kyoto School' as the archetype of infinite-dimensional integrable systems \cite{MiwaJimboDate2000,Dickey2003}.
\citet{Kako1976:KP} derived three types of two-dimensional generalizations of Washimi and Taniuti's result, including the KP equation.
% These equations are integrable, and solitary waves are called soliton.
Diverse directions of generalizations have been also studied; for example, variations of the KdV equation including
finite ion temperature \cite{Tappert1972,Tagare1973:KdV},
multi-ions \cite{Tran-Hirt1974},
and dust plasma \cite{Tiwari2006};
as well as variations of the KP equation including
multi-ions \cite{UrRehman2013},
dust plasma \cite{Duan2002},
and multi-temperature \cite{LinDuan2005:KP,Gill2006}.
The modifications to include third-order nonlinear terms were proposed by considering trapped electrons \cite{Schamel1973:mKdV,OKeir1997}.
%mKdV: three-waves interaction,\cite{Konno1974:mKdV} multi-ions,\cite{Kalita1990}
Effects of higher order terms in the reductive perturbation method have been also widely studied 
%\cite{*[][. See also Ref.~\cite{Tiwari2006} and references therein.] Ichikawa1977:dress1,Ichikawa1977:dress2,Konno1977:dress%,Sugimoto1977,Tiwari2006}.
(see, e.g., Refs.~\cite{Konno1977:dress,Ichikawa1977:dress1,Tiwari2006} and references therein).
The higher order perturbations are called clouds, and the solitons are called dressed solitons.

%% Vorticity %%%%%%%%%%%%
In this paper, we explore a new direction of generalization---we introduce a \emph{vorticity} to the system, 
and delineate a fundamental change of dynamics brought about by the vortex.
The KP hierarchy is characterized not only by small amplitudes but also by irrotational (zero-vorticity) velocity fields
(see Section \ref{sec:KP_zeroV1}).
We may view the ordered system of solitons as a singular submanifold (leaf) embedded in a larger phase space
of finite-vorticity perturbations \cite{YoshidaMorrison2016:phantom}.
The departure from the zero-vorticity leaf will produce complexity and, finally, generate turbulence.
The aim of this study is to probe into the `neighborhood' of the KP hierarchy and elucidate how chaos starts to develop.

We organize this paper as follows. 
In Section \ref{sec:KP}, we show that the KP equation is derived by eliminating vorticity at every order of the reductive perturbation.
We also show that the reductive perturbation succeeds only if the entropy is homogeneous;
hence the baroclinic effect, a creation mechanism of vorticity, must be absent (\ref{sec:KP_baroc}).
In Section \ref{sec:KPY}, we introduce a new ordering of velocity field in order to formulate a finite-vorticity system.
The new system is composed of a generalized 
three-dimensional 
KP equation and a two-dimensional vorticity equation.
The former describes `scattering' of vortex-free waves by ambient vortexes that are determined by the latter.
We say that the vortexes are `ambient' because they do not receive reciprocal reactions from the waves.
In Section \ref{sec:Painleve}, we invoke the Painlev{\'e} test to study whether the new system is integrable or not.
The result is negative. 
By this analysis, we elucidate that the scattering by the vorticity introduces an essential three-dimensionality in the wave fields,
by which the integrability condition (in the sense of the Painlev\'e test) is broken.
In Section \ref{sec:numerical}, we perform numerical simulations to visualize how chaos occurs.
Section \ref{sec:conclusion} concludes our investigations.

%%%%%%%%%%%%%%%%%%%%%%%%%%%%%%%%%%%%%%%%%%%%%%%%%%%%%%%%%%%%%%%%
%%%%%%%%%%%%%%%%%%%%%%%%%%%%%%%%%%%%%%%%%%%%%%%%%%%%%%%%%%%%%%%%
%%%%%%%%%%%%%%%%%%%%%%%%%%%%%%%%%%%%%%%%%%%%%%%%%%%%%%%%%%%%%%%%
\section{Reductive perturbation method for Kadomtsev--Petviashvili equation and vorticity}\label{sec:KP}
%%%%%%%%%%%%%%%%%%%%%%%%%%%%%%%%%%%%%%%%%%%%%%%%%%%%%%%%%%%%%%%%
%%%%%%%%%%%%%%%%%%%%%%%%%%%%%%%%%%%%%%%%%%%%%%%%%%%%%%%%%%%%%%%%
%%%%%%%%%%%%%%%%%%%%%%%%%%%%%%%%%%%%%%%%%%%%%%%%%%%%%%%%%%%%%%%%

%%%%%%%%%%%%%%%%%%%%%%%%%%%%%%%%%%%%%%%%%%%%%%%%%%%%%%%%%%%%%%%%
\subsection{Kadomtsev--Petviashvili equation}\label{sec:KP_deriv}
%%%%%%%%%%%%%%%%%%%%%%%%%%%%%%%%%%%%%%%%%%%%%%%%%%%%%%%%%%%%%%%%
We start by remembering the derivation of the KP equation by the \emph{reductive perturbation method} \cite{Washimi1966:KdV,Kako1976:KP,Kadomtsev-Petviashvili}.
The basic equations for nonlinear IAWs are expressed as
\begin{align}
    \label{eq:cont}  & \pdif{n}{t} + \nabla \cdot (n \bm u) = 0, \\
    \label{eq:mom}  & \pdif{\bm u}{t} + (\bm u \cdot \nabla) \bm u = \nabla \phi, \\
    \label{eq:Pois}  & - \lap \phi = n - \rme^\phi,
\end{align}
where $n$ is the ion number density, $\bm u =(u,v,w)^\top$ is the ion velocity, $\phi$ is the electrostatic potential, and $\lap$ is the Laplacian.
We consider cold ions and adiabatic electrons with a constant temperature $T_\rme$.
The variables are normalized as followings: 
the density $n$ by a representative density $n_0$, 
the velocity $\bm u$ by the ion sound speed $c_\rms = \sqrt{T_\rme/m_\rmi}$ (where $m_\rmi$ is the ion mass),
the electrostatic potential $\phi$ by the characteristic potential $T_\rme/e$, 
the coordinate variable $\bm x$ by the Debye length $\sqrt{\varepsilon_0 T_\rme/n_0 e^2}$,
and the time variable $t$ by the ion plasma frequency $\sqrt{n_0 e^2/\varepsilon_0 m_\rmi}$.

We consider IAWs propagating in two-dimensional space $(x,y)$.
The extension to three-dimensional space $(x,y,z)$ will be discussed later.
We assume that waves propagate primarily in the direction of $x$, and introduce a set of stretched variables
\begin{equation}\label{eq:rp_xyt}
    \tilde x = \epsilon (x-t), \qquad
    \tilde y = \epsilon^2 y, \qquad
    \tilde t = \epsilon^3 t,
\end{equation}
with a small parameter $\epsilon$.
The dependent variables $n$, $\phi$, $u$, and $v$ are expanded as
\begin{equation}\label{eq:rp}
\left\{ \begin{aligned}
    n &= 1 + \epsilon^2 n_1 + \epsilon^4 n_2 + \cdots, \\
    \phi &= 0 + \epsilon^2 \phi_1 + \epsilon^4 \phi_2 + \cdots, \\
    u &= 0 + \epsilon^2 u_1 + \epsilon^4 u_2 + \cdots, \\
    v &= 0 + \epsilon^3 v_1 + \epsilon^5 v_2 + \cdots.
\end{aligned} \right.
\end{equation}
%From the terms of orders $\epsilon^2$ and $\epsilon^3$, we obtain $n_1 = \phi_1$.
From the terms of orders $\epsilon^2$ and $\epsilon^3$, we obtain $n_1 = \phi_1$ and $\pd n_1/\pd\tilde x = \pd u_1/\pd\tilde x = \pd \phi_1/\pd\tilde x$.
Assuming the boundary conditions $n_1, \phi_1, u_1 \to 0$ $(x \to \pm \infty)$,  we put
\begin{equation}\label{eq:KP.e1}
    n_1 = u_1 = \phi_1.
\end{equation}
From the terms of order $\epsilon^4$, we obtain 
\begin{equation}\label{eq:KP.e2y}
    \pdif{v_1}{\tilde x} = \pdif{\phi_1}{\tilde y}
\end{equation}
and $n_2 = \phi_2 + \phi_1^2/2 - \pd^2 \phi_1 / \pd {\tilde x}^2$.
From the terms of order $\epsilon^5$, we obtain the two-dimensional KP equation:
\begin{equation}\label{eq:KP_2D}
    \pdif{}{\tilde x} \left( \pdif{\phi_1}{\tilde t} +\phi_1 \pdif{\phi_1}{\tilde x} + \frac12 \pdif[3]{\phi_1}{\tilde x} \right) + \frac12 \pdif[2]{\phi_1}{\tilde y} = 0.
\end{equation}

In three-dimensional space $(x,y,z)$, we introduce a stretched variable $\tilde z$ and expand the velocity $w$ as
\begin{equation}\label{eq:rp_z}
    \tilde z = \epsilon^2 z, \qquad
    w = 0 + \epsilon^3 w_1 + \epsilon^5 w_2 + \cdots.
\end{equation}
From the terms of order $\epsilon^4$, we obtain the relation
\begin{equation}\label{eq:KP.e2z}
    \pdif{w_1}{\tilde x} = \pdif{\phi_1}{\tilde z} ,
\end{equation}
which parallels equation \eqref{eq:KP.e2y}.
The terms of order $\epsilon^5$ lead to the three-dimensional KP equation
\begin{equation}\label{eq:KP_3D}
    \pdif{}{\tilde x} \left( \pdif{\phi_1}{\tilde t} +\phi_1 \pdif{\phi_1}{\tilde x} + \frac12 \pdif[3]{\phi_1}{\tilde x} \right) + \frac12 \lap_\perp \phi_1 = 0,
\end{equation}
where $\lap_\perp = (\pd/\pd {\tilde y})^2 + (\pd/\pd {\tilde z})^2$.
% is the Laplacian perpendicular to the primal propagating direction ($x$).

% To be precise, the KP equation [Eq.~\eqref{eq:KP_2D} or \eqref{eq:KP_3D}] derived in this section is called KP-II equation.

%%%%%%%%%%%%%%%%%%%%%%%%%%%%%%%%%%%%%%%%%%%%%%%%%%%%%%%%%%%%%%%%
\subsection{Vorticity of Kadomtsev--Petviashvili equation}\label{sec:KP_zeroV1}
%%%%%%%%%%%%%%%%%%%%%%%%%%%%%%%%%%%%%%%%%%%%%%%%%%%%%%%%%%%%%%%%
% We find that the vorticity of the KP equation is zero.
Strikingly absent in the KP system is the vorticity.
The $z$-component of the vorticity $\omega_z = \pd v/\pd x - \pd u/\pd y$ is expanded as
\begin{equation}\label{eq:omega_ord_z}
    \omega_z = \epsilon^4 \left( \pdif{v_1}{\tilde x} - \pdif{u_1}{\tilde y} \right)
        + \epsilon^6 \left( \pdif{v_2}{\tilde x} - \pdif{u_2}{\tilde y} \right) + \cdots .
\end{equation}
The leading-order term turns out to be zero from equations \eqref{eq:KP.e1} and \eqref{eq:KP.e2y}.
The same ordering applies to the $y$-component of the vorticity $\omega_y = \pd u/\pd z - \pd w/\pd x$;
hence the leading-order vorticity vanishes by the demand of equations \eqref{eq:KP.e1} and \eqref{eq:KP.e2z}.
The $x$-component of the vorticity $\omega_x = \pd w/\pd y - \pd v/\pd z$ is expanded as
\begin{equation}\label{eq:omega_ord_x}
    \omega_x = \epsilon^5 \left( \pdif{w_1}{\tilde y} - \pdif{v_1}{\tilde z} \right)
        + \epsilon^7 \left( \pdif{w_2}{\tilde y} - \pdif{v_2}{\tilde z} \right) + \cdots.
\end{equation}
The ordering is slightly different from those of $\omega_y$ and $\omega_z$.
However, the leading-order term is forced to vanish by equations \eqref{eq:KP.e1}, \eqref{eq:KP.e2y}, \eqref{eq:KP.e2z},
and the boundary condition $\pd w_1/\pd \tilde y - \pd v_1/\pd \tilde z \to 0$ $(x \to \pm \infty)$.
Thus, the KP system is vortex-free.

In \ref{sec:KP_baroc}, we discuss a finite-temperature system, in which an inhomogeneous entropy 
yields a baroclinic effect, a mechanism creating vorticity.
Then, the aforementioned ordering fails to formulate a reductive perturbation, 
implying that the KP system is fundamentally incapable to host a vorticity. 

% Now we try to introduce vortexes.
% We start by investigating whether by some variations used in the existing works, 
% higher order perturbation terms and finite ion temperature, can produce vorticities.
% We examine whether higher order perturbation terms have vorticity in the next section 
% and examine baroclinic effect provided by finite ion temperature in \ref{sec:KP_baroc}.
% The results are both negative and motivate us to modify the reductive perturbation method (Section \ref{sec:KPY}).

%%%%%%%%%%%%%%%%%%%%%%%%%%%%%%%%%%%%%%%%%%%%%%%%%%%%%%%%%%%%%%%%
\subsection{Vorticity of general order}\label{sec:KP_zeroV2}
%%%%%%%%%%%%%%%%%%%%%%%%%%%%%%%%%%%%%%%%%%%%%%%%%%%%%%%%%%%%%%%%
The absence of vorticity is not only at the order of KP equation, but also at all orders of perturbations.
Let us examine higher order equations.
The second-order equation is linear with respect to the second-order variables,
and includes an inhomogeneous term depending on $\phi_1$ \cite{Ichikawa1977:dress1,Konno1977:dress,Tiwari2006}.
The higher order perturbations are called clouds surrounding the core, i.e., the first-order perturbation;
a soliton with clouds is called a dressed soliton.
As proved in the previous section, the core is vortex-free.
Moreover, we find that all higher order clouds are vortex-free.
% One can verify, by direct calculations, that higher order vorticities are zero. 
% However, calculations of higher order terms are rather elaborating.
% We show it by a simpler procedure.

The vorticity equation is derived by taking the curl $\nabla \times {}$ of the equation of motion \eqref{eq:mom}:
\begin{equation}\label{eq:vort}
    \pdif{\bm\omega}{t} = \nabla \times (\bm u \times \bm \omega).
\end{equation}
% There are no source terms since we consider only the potential force $\nabla\phi$.
After the Galilean boost in the $x$-direction (see equation \eqref{eq:rp_xyt}), 
equation \eqref{eq:vort} reads as
\begin{equation}\label{eq:vort_boost}
    \pdif{\bm\omega}{t} - \pdif{\bm\omega}{x} = \nabla \times (\bm u \times \bm \omega).
\end{equation}
Inserting the expressions \eqref{eq:rp_xyt} and \eqref{eq:rp}, and using 
equations \eqref{eq:rp_z}, \eqref{eq:omega_ord_z}, and \eqref{eq:omega_ord_x},
let us see the orders of each term in equation \eqref{eq:vort_boost};
in the $x$-component
\begin{equation}\label{eq:vort_boost_x}
    \pdif{\omega_x}{t} - \pdif{\omega_x}{x} = \pdif{}{y} (u \omega_y - v \omega_x) - \pdif{}{z} (w \omega_x - u \omega_z),
\end{equation}
the order of the second term on the left-hand-side is $\epsilon^6$, while all other terms are of order $\epsilon^8$.
Thus, the lowest order vorticity $\omega_{x1}$ must satisfy $\pd \omega_{x1}/\pd x =0$.
By the boundary condition $\omega_{x1} \to 0$ $(x \to \pm \infty)$, therefore,
we obtain $\omega_{x1}=0$.
In the $y$- and $z$-components of equation \eqref{eq:vort_boost}, the orders of all terms decrease by $\epsilon^1$;
hence we obtain $\omega_{y1}=\omega_{z1}=0$ under the boundary conditions $\omega_{y1},\omega_{z1} \to 0$ $(x \to \pm \infty)$.
% same as in the case of $\omega_{x1}$ 

%Next, in the $y$-component of equation \eqref{eq:vort_boost},
%\begin{equation}\label{eq:vort_boost_y}
%    \pdif{\omega_y}{t} - \pdif{\omega_y}{x} = \pdif{}{z} (v \omega_z - w \omega_y) - \pdif{}{x} (u \omega_y - v \omega_x),
%\end{equation}
%the second term of left-hand-side has order $\epsilon^5$, and the lowest order of other terms is $\epsilon^7$.
%Same evaluation is valid for $\omega_z$, hence we obtain $\omega_{z1}=0$ with the boundary condition $\omega_{z1} \to 0$ $(x \to \pm \infty)$.

Eliminating $\bm\omega_1$, 
equation \eqref{eq:vort_boost} reads as the equation for $\bm\omega_2$ (the second-order part of $\bm\omega$).
By the same argument of ordering, we obtain $\bm\omega_2=0$, and 
equation \eqref{eq:vort_boost} dominates the next order.
Continuing the induction, we conclude that  $\bm\omega_j=0$ for all order $j$.

%%%%%%%%%%%%%%%%%%%%%%%%%%%%%%%%%%%%%%%%%%%%%%%%%%%%%%%%%%%%%%%%
%%%%%%%%%%%%%%%%%%%%%%%%%%%%%%%%%%%%%%%%%%%%%%%%%%%%%%%%%%%%%%%%
%%%%%%%%%%%%%%%%%%%%%%%%%%%%%%%%%%%%%%%%%%%%%%%%%%%%%%%%%%%%%%%%
\section{Generalized system with finite vorticity}\label{sec:KPY}

We can formulate a generalized system with a finite vorticity by introducing a velocity $\bm v_0 = (0, v_0, w_0)^\top$ of order $\epsilon^1$ in the $y$- and $z$-directions:
\begin{equation}\label{eq:rp_KPY}
\left\{ \begin{aligned}
    v &= \epsilon v_0 + \epsilon^3 v_1 + \epsilon^5 v_2 + \cdots, \\
    w &= \epsilon w_0 + \epsilon^3 w_1 + \epsilon^5 w_2 + \cdots.
\end{aligned} \right.
\end{equation}
We assume that the additional velocity is homogeneous in the $x$-direction ($\pd \bm v_0/\pd x = \bm 0$) 
and incompressible ($\nabla \cdot \bm v_0 = 0$).
By the two-dimensionality and incompressibility, we may write $\bm v_0$ in a Clebsch form \cite{Yoshida2009:Clebsch}
\begin{equation}\label{eq:v0_Clebsch}
    \bm v_0 = \nabla_\perp \psi (\tilde y, \tilde z) \times \bm e_x,
\end{equation}
where $\bm e_x$ is the unit vector in the $x$-direction, $\psi$ is the stream function, and $\nabla_\perp = (0,\pd/\pd\tilde y,\pd/\pd\tilde z)$.

From the lowest order terms, we obtain the conventional two-dimensional Euler vorticity equation
\begin{equation}\label{eq:2DEV}
    \pdif{}{\tilde t} \lap_\perp \psi + [\lap \psi, \psi] = 0,
\end{equation}
where $\lap_\perp \psi$ denotes the vorticity and $[f,g] = \bm e_x \cdot (\nabla_\perp f \times \nabla_\perp g) = (\pd f/\pd \tilde y)(\pd g/\pd \tilde z) - (\pd f/\pd \tilde z)(\pd g/\pd \tilde y)$.
From the terms of order $\epsilon^5$, we obtain a three-dimensional wave equation:
\begin{equation}\label{eq:KPY}
    \pdif{}{\tilde x} \left( \pdif{\phi_1}{\tilde t} +\phi_1 \pdif{\phi_1}{\tilde x} + \frac12 \pdif[3]{\phi_1}{\tilde x} + [\phi_1, \psi] \right) + \frac12 \lap_\perp \phi_1 = 0.
\end{equation}
For the convenience, we call the system of equations \eqref{eq:2DEV} and \eqref{eq:KPY} Kadomtsev--Petviashvili--Yoshida (KPY) equations
and compare it with the KP equation.
In what follows, the subscript ${}_1$ and tilde $\tilde{\hphantom{x}}$ will be omitted for simplicity.

The vortex field $\psi$ ($\lap_\perp\psi$ is the vorticity)
affects the wave field $\phi$ through equation \eqref{eq:KPY} 
while it does not receive any reciprocal reaction from $\phi$
(notice that the vortex equation \eqref{eq:2DEV} is independent of $\phi$).
This is because the order of $\psi$ is lower than that of $\phi$.
We may treat $\psi$ as an `ambient' field for $\phi$.

% The similar relation is found in a study of Alfv\'en waves in the incompressible ideal magnetohydrodynamic model \cite{Hirota2005}.
% In their model, the velocity and magnetic field $v_x \pm B_x$ affect 
% the behaviors of vorticity and current $\omega_x \pm j_x$ (the current field is the rotation of the magnetic field: $\bm j = \nabla \times \bm B$), 
% but the former is not affected by the latter.

In the rest of paper, we assume that the vortex field $\psi$ is stationary. 
For example, $\psi = a \sin (k y) - a \cos (k z)$ is a stationary solution of the Euler equation \eqref{eq:2DEV}, 
which satisfies $\lap_\perp \psi \propto \psi$.
Then, only \eqref{eq:KPY}, to be called the KPY equation, will be the target of analysis.

%%%%%%%%%%%%%%%%%%%%%%%%%%%%%%%%%%%%%%%%%%%%%%%%%%%%%%%%%%%%%%%%
%%%%%%%%%%%%%%%%%%%%%%%%%%%%%%%%%%%%%%%%%%%%%%%%%%%%%%%%%%%%%%%%
%%%%%%%%%%%%%%%%%%%%%%%%%%%%%%%%%%%%%%%%%%%%%%%%%%%%%%%%%%%%%%%%
\section{Painlev\'e analysis}\label{sec:Painleve}
%%%%%%%%%%%%%%%%%%%%%%%%%%%%%%%%%%%%%%%%%%%%%%%%%%%%%%%%%%%%%%%%
%%%%%%%%%%%%%%%%%%%%%%%%%%%%%%%%%%%%%%%%%%%%%%%%%%%%%%%%%%%%%%%%
%%%%%%%%%%%%%%%%%%%%%%%%%%%%%%%%%%%%%%%%%%%%%%%%%%%%%%%%%%%%%%%%
Here we put the KPY equation to the Painlev\'e test.
Let us begin by reviewing the method briefly.
According to Weiss, Tabor, and Carnevale (WTC) \cite{WTC_Painleve}, 
a nonlinear partial differential equation (PDE) is said to have the \emph{Painlev\'e property} 
when the solutions are represented in terms of a Laurent series 
in a neighborhood of a movable singularity manifold
(to be identified as a set of points satisfying $\varphi(x,y,z,t) = 0$);
assuming that the solution $\phi(x,y,z,t)$ of PDE can be written,
in the neighborhood of the singularity manifold, as
\begin{equation}\label{eq:Laurent}
    \phi(x,y,z,t) = \chi^{-q} \sum_{j=0}^\infty \phi_j(x,y,z,t) \chi^j
\end{equation}
with analytic functions $\phi_j$ and an expansion function $\chi$ which vanishes as $\varphi\rightarrow0$,
the Painlev\'e property is tested by
% (i) substituting the Laurent series \eqref{eq:Laurent} into the PDE, and (ii) 
verifying that $q$ is a positive integer and that all $\phi_j$'s can be determined consistently.
% This procedure is called the \emph{Painlev\'e test}.
There are various choices of the expansion function $\chi$.
The most natural one is $\chi = \varphi$, which is used by WTC \cite{WTC_Painleve}.
However, this choice makes $\phi_j$ complicated.
% The simplest choose for the test is $\varphi(x,y,z,t) = x - f(y,z,t)$, which is called Kruskal assumption.\cite{WTC_Painleve,Jimbo1982}
\citet{Conte1989} has shown that the best choice for reducing the calculation without any constraints on $\varphi$ is
\begin{equation}\label{eq:Conte}
    \chi = \left( \frac{\varphi_x}{\varphi} - \frac{\varphi_{xx}}{2\varphi_x} \right)^{-1},
\end{equation}
where subscripts denote derivatives, e.g., $\varphi_x=\pd\varphi/\pd x$.
(A constraint $\varphi_x = 0$ is required, however, this is also necessary for WTC and other choices.)
For details of the Painlev\'e analysis for PDEs, see, e.g., Ref.~\cite{Musette1999}.

The Painlev\'e property is considered to be equivalent to the integrability of a PDE, 
and many integrable equations (e.g., the Burgers equation, the KdV equation, and the two-dimensional KP equation) pass the Painlev\'e test \cite{WTC_Painleve}.
However, the three-dimensional KP equation does not pass the test \cite{Brugarino1994,Ruan1999,Xu2004}, and it is not integrable
(see, e.g., Ref.~\cite{Ma2011:3+1KP} for another explanation of its non-integrability).
% \citet{Brugarino1994} have shown with WTC choice $\chi=\varphi$ and \citet{Ruan1999} have shown with Conte's choice \eqref{eq:Conte}.
Since the additional vortex field is perpendicular to the primal direction of propagation, it brings about an essential three-dimensionality.
Thus, it is expected that the KPY equation \eqref{eq:KPY} is not integrable.

Now we execute the Painlev\'e test for the KPY equation \eqref{eq:KPY} with Conte's choice \eqref{eq:Conte} 
and show that the KPY equation passes the Painlev\'e test only under some special conditions.
In order to elucidate when the equation is integrable or not, we use a generalized form
\begin{equation}\label{eq:KPYm}
    \phi_{xt} + (\phi\phi_x)_x + \alpha \phi_{4x} + \beta \phi_{yy} + \gamma \phi_{zz} 
    + a(y,z) \phi_{xy} + b(y,z) \phi_{xz} = 0,
\end{equation}
where $\alpha, \beta, \gamma$ are constants with $\alpha \neq 0$ and $a,b$ are functions of $y,z$.
In the original form \eqref{eq:KPY}, coefficients are chosen as $\alpha=1$, $\beta=\gamma=1/2$, $a(y,z) = \pd\psi/\pd z$, and $b(y,z) = - \pd\psi/\pd y$. 

The leading-order analysis (substituting $\phi=\phi_0\chi^{-q}$ and comparing leading-order terms) 
determines the values of $q$ and $\phi_0$ as $q=2$ and $\phi_0 = -12 \alpha$.
From general order terms, we obtain recursion relations
\begin{equation}\label{eq:rec}
    (j+1)(j-4)(j-5)(j-6) \alpha \phi_j = F_j(\phi_0,\ldots, \phi_{j-1})
\end{equation}
for $j=1,2,\ldots{}$, where $F_j$'s are complicated functions of $\phi_0,\ldots,\phi_{j-1}$, $\varphi$, and their derivatives.
$\phi_1$, $\phi_2$, and $\phi_3$ are determined by equation \eqref{eq:rec}, 
and $F_4 = F_5 = F_6 = 0$ is required for consistency (called a resonance condition).
%When the condition is satisfied, 
When the condition puts no constraint on $\varphi$, 
we can determine all $\phi_j$'s cotnsistently with arbitrary functions $\phi_4$, $\phi_5$, and $\phi_6$.
The two-dimensional KP equation \eqref{eq:KP_2D} satisfies the resonance condition,
however, the three-dimensional KP equation \eqref{eq:KP_3D} and KPY equation \eqref{eq:KPY} do not satisfy that. %they are not Painlev\'e integrable.
In the latter case, we find $F_4=F_5=0$ and $F_6 \neq 0$.
For the KPY equation \eqref{eq:KPYm}, $F_6$ is expressed as
\begin{align}
    F_6 &= \beta\gamma E_1(\varphi) + \beta E_2(a_y, a_{yy}, b_y, b_{yy}; \varphi)
        + \gamma E_3(a_z, a_{zz}, b_z, b_{zz}; \varphi),
\\
    E_1 &= \frac{4}{\varphi_x^4}
        \bigl[ \varphi_x^2 (\varphi_{yy} \varphi_{yz} - \varphi_{yz}^2)
            + 2 \varphi_x \varphi_y (\varphi_{xz} \varphi_{yz} - \varphi_{xy} \varphi_{zz}) + \circlearrowleft_{x,y,z}
        \bigr],
\\
    E_2 &= \frac{1}{\varphi_x^3} \bigl[
            - a_{yy} \varphi_x^2 \varphi_y - b_{yy} \varphi_x^2 \varphi_z 
            - 2 a_y \varphi_x^2 \varphi_{yy} - 2 b_y \varphi_x^2 \varphi_{yz}
\nonumber \\
            &\hspace{.42in} + 2 (2  a_y +  b_y ) \varphi_x \varphi_z \varphi_{xy}
            + 2 b_y \varphi_x \varphi_y \varphi_{xz} 
            - 2 b_y \varphi_{xx} \varphi_y \varphi_z -2 a_y \varphi_{xx} \varphi_y^2
        \bigr],
\\
    E_3 &= \frac{1}{\varphi_x^3} \bigl[
            - a_{zz} \varphi_x^2 \varphi_y - b_{zz} \varphi_x^2 \varphi_z 
            - 2 a_z \varphi_x^2 \varphi_{yz} - 2 b_z \varphi_x^2 \varphi_{zz}
\nonumber \\
            &\hspace{.42in} + 2 (a_z + 2 b_z) \varphi_x \varphi_y \varphi_{xz}
            + 2 a_z \varphi_x \varphi_z \varphi_{xy}
            - 2 a_z \varphi_{xx} \varphi_y \varphi_z - 2 b_z \varphi_{xx} \varphi_z^2
        \bigr].
\end{align}
where $\circlearrowleft_{x,y,z}$ denotes the summation over cyclic permutation of $x,y,z$.
$E_1$ is $F_6$ of the three-dimensional KP equation \cite{Brugarino1994}.

The resonance condition $F_6 = 0$ is satisfied only in the following special cases:
(i) $\beta = \gamma = 0$; (ii) $\gamma = 0$, $a_y = b_y = 0$; (iii) $\beta = 0$, $a_z = b_z = 0$.
In the case (i), the KPY equation \eqref{eq:KPYm} is reduced to the one-dimensional $(x)$ KdV equation with advection terms in the independent directions ($y,z$).
In the case (ii), the KPY equation is reduced to the two-dimensional $(x,y)$ KP equation with Galilean boosting (whose speed is homogeneous in $x$- and $y$-directions).
The case (iii) is same as the case (ii), with $y$ and $z$ exchanged.
The cases (ii) and (iii) are consistent with integrable conditions of generalized variable-coefficient two-dimensional KP equations, see, e.g., Ref.~\cite{Tian2012}.

From the above result, we find that the KPY equation is integrable only if it can be reduced to a lower dimensional integrable system
(the one-dimensional KdV equation or the two-dimensional KP equation).
In the next section, we demonstrate chaotic behaviors of IAW by numerical solutions of the KPY equation.

%%%%%%%%%%%%%%%%%%%%%%%%%%%%%%%%%%%%%%%%%%%%%%%%%%%%%%%%%%%%%%%%
%%%%%%%%%%%%%%%%%%%%%%%%%%%%%%%%%%%%%%%%%%%%%%%%%%%%%%%%%%%%%%%%
%%%%%%%%%%%%%%%%%%%%%%%%%%%%%%%%%%%%%%%%%%%%%%%%%%%%%%%%%%%%%%%%
\section{Numerical analysis}\label{sec:numerical}
%%%%%%%%%%%%%%%%%%%%%%%%%%%%%%%%%%%%%%%%%%%%%%%%%%%%%%%%%%%%%%%%
%%%%%%%%%%%%%%%%%%%%%%%%%%%%%%%%%%%%%%%%%%%%%%%%%%%%%%%%%%%%%%%%
%%%%%%%%%%%%%%%%%%%%%%%%%%%%%%%%%%%%%%%%%%%%%%%%%%%%%%%%%%%%%%%%
\subsection{Settings of numerical simulation}
We perform numerical simulation by the following setting.
We consider a domain $(x,y,z) \in [0,20] \times [-5,5] \times [-5,5]$ with the periodic boundary condition.
The KPY equation \eqref{eq:KPY} is solved in the following splitting form:
\begin{align}
    \pdif{\phi}{t} &= \mathcal{L} \phi + \mathcal{N}(\phi), \\
    \calL \phi &= - \frac{1}{2} \pdif[3]{\phi}{x} - \frac{1}{2} \pd_x^{-1} \lap_\perp \phi, \\
    \calN(\phi) &= - \frac{1}{2} \pdif{}{x} (\phi^2) - [\phi, \psi],
\end{align}
with the second-order Strang splitting method \cite{Strang1968}
\begin{equation}\label{eq:Strang}
    \phi_{n+1} = \exp \left( \frac{h}{2} \calL \right) \exp(h \calN) \exp\left( \frac{h}{2} \calL \right) \phi_n,
\end{equation}
where $h$ is the time step size.
Each factor on the right-hand side of equation \eqref{eq:Strang} must be approximated by a second or higher order scheme.
Here, the linear evolution $\exp(h\calL)$ is solved implicitly with the Fourier transformation,
while the nonlinear evolution $\exp(h \calN)$ is approximated by the second-order Runge--Kutta method with finite-difference approximations.
The anti-derivative $\pd_x^{-1}$ in the linear operator $\calL$ can be calculated by the Fourier multiplier $-\rmi/k_x$.
In order to regularize the singularity at $k_x=0$, this multiplier is modified as $-\rmi/(k_x + \rmi \delta)$,
with a small real number $\delta$ (we use the machine epsilon of double precision floating point number $2^{-52} \sim 2.2 \times 10^{-16}$) \cite{Klein2007,Klein2011,Einkemmer2015:KP}.
% It should be noted that, when the sign before $\pd_x^{-1}\Delta_\perp\phi$ is reversed, the sign of $+\rmi \delta$ must be also reversed.

We give an initial condition by modifying 
the single soliton solution of the two-dimensional KP equation:
\begin{equation}
    \phi_0(x,y,z) = 3 A \operatorname{sech}^2 \left[ \sqrt{\frac{A}{2}} (x - B y - C) \right],
\end{equation}
where $A$, $B$, and $C$ are arbitrary constants (we choose $A=1$, $B=2$, and $C=10$).
% We assume that the initial state of $\phi$ is homogeneous in the $z$-direction.
To put the waves in the periodic domain, $x$ and $y$ are, respectively, modulo 20 and 10 (box sizes).
Furthermore, $\phi$ must satisfy
\begin{equation}\label{eq:constraint}
\int \Delta_\perp \phi \, \rmd x = 0,
\end{equation}
which is derived by integrating the KPY equation \eqref{eq:KPY} in the $x$-direction.
By subtracting the Fourier components with $k_x=0$ ($k_x$ is the wave number in the $x$ direction),
excepting the $k_x=k_y=k_z=0$ component, from $\phi_0(x,y,z)$, we obtain the hoped-for initial condition.
%It is known that the solution to the 
The KP equation (i.e., $\psi=0$), starting from this initial condition, propagates with conserving the wave shape.

As described at the end of Section \ref{sec:KPY}, we assume that $\psi$ is stationary.
We use 
\begin{equation}
    \psi(y,z) = a \sin \left( \frac{2\pi \kappa y}{L} \right) - a \cos \left( \frac{2\pi\kappa z}{L} \right),
\end{equation}
where $L$ is the length of the domain in the $y$- and $z$-directions ($L=10$).
$a$ and $\kappa$ denote the intensity and the wavenumber of the vortex field.
We change their values and observe how line-solitons are `scattered' by  the vortex.

%%%%%%%%%%%%%%%%%%%%%%%%%%%%%%%%%%%%%%%%%%%%%%%%%%%%%%%%%%%%%%%%
%%%%%%%%%%%%%%%%%%%%%%%%%%%%%%%%%%%%%%%%%%%%%%%%%%%%%%%%%%%%%%%%
%%%%%%%%%%%%%%%%%%%%%%%%%%%%%%%%%%%%%%%%%%%%%%%%%%%%%%%%%%%%%%%%
\begin{figure}[t]
\begin{minipage}{.475\hsize}\centering
\includegraphics[scale=.8]{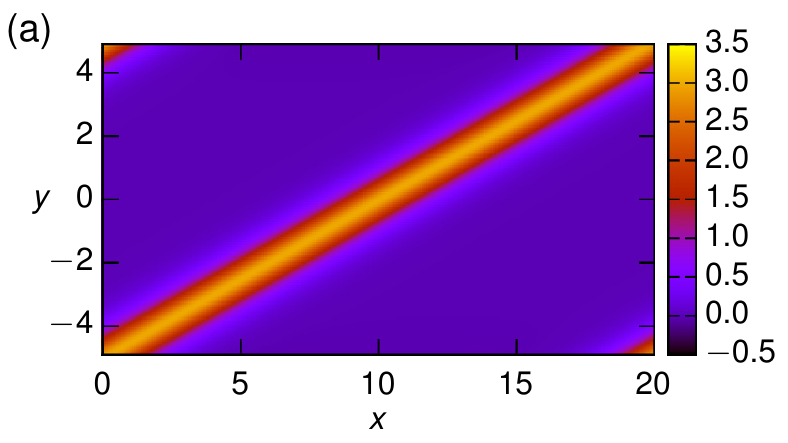}
\end{minipage}
\begin{minipage}{.475\hsize}\centering
\includegraphics[scale=.8]{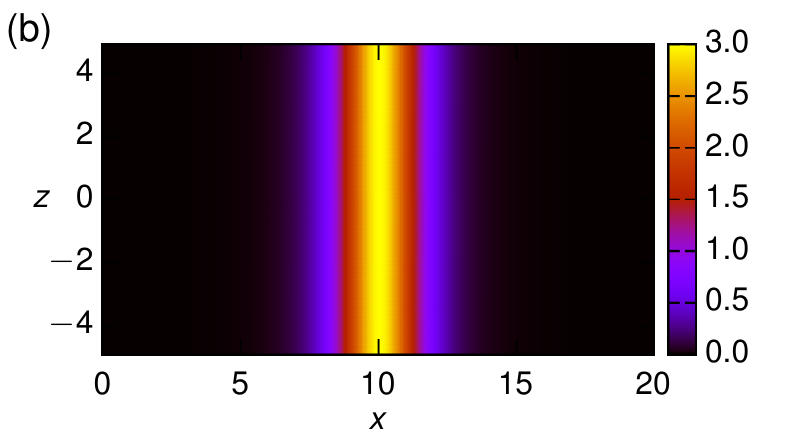}
\end{minipage}
\caption{Cross-sections of $\phi$ at $t=0$ (a line-soliton with the periodic boundary condition, homogeneous in the $z$-direction). 
(a) $(z=0)$-cross-section, (b) $(y=0)$-cross-section.}
\label{fig:line-soliton}
\end{figure}
%%%%%%%%%%%%%%%%%%%%%%%%%%%%%%%%%%%%%%%%%%%%%%%%%%%%%%%%%%%%%%%%
%%%%%%%%%%%%%%%%%%%%%%%%%%%%%%%%%%%%%%%%%%%%%%%%%%%%%%%%%%%%%%%%
%%%%%%%%%%%%%%%%%%%%%%%%%%%%%%%%%%%%%%%%%%%%%%%%%%%%%%%%%%%%%%%%

%%%%%%%%%%%%%%%%%%%%%%%%%%%%%%%%%%%%%%%%%%%%%%%%%%%%%%%%%%%%%%%%
\subsection{Numerical results}\label{sec:result}
\begin{figure}[t]
\begin{minipage}{.475\hsize}\centering
\includegraphics[scale=.8]{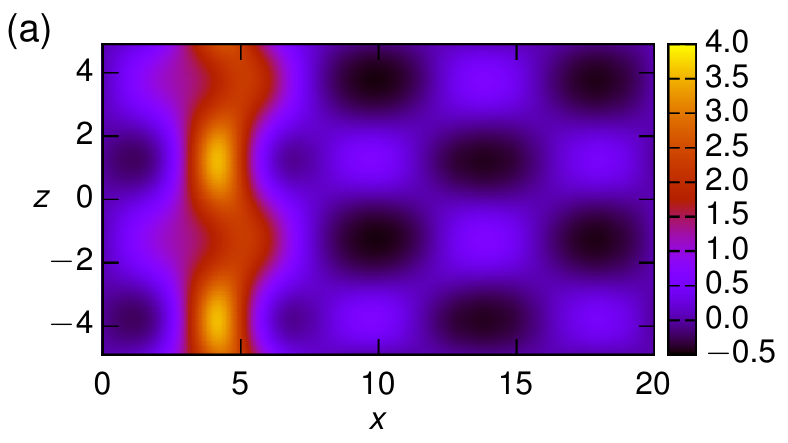}
\end{minipage}
\begin{minipage}{.475\hsize}\centering
\includegraphics[scale=.8]{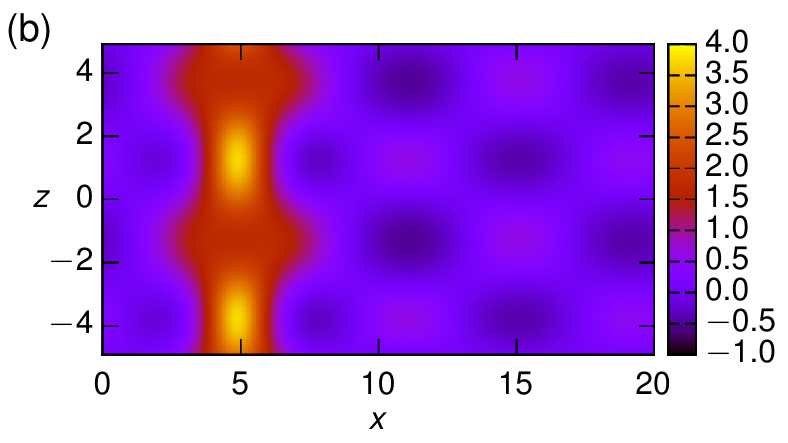}
\end{minipage}\\
\begin{minipage}{.475\hsize}\centering
\includegraphics[scale=.8]{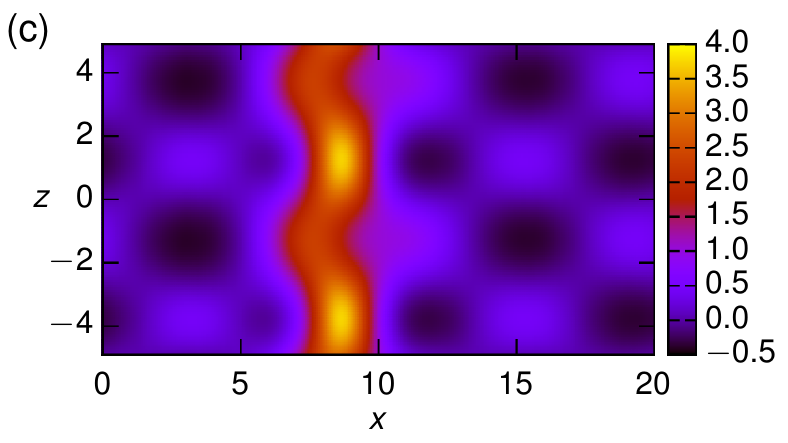}
\end{minipage}
\begin{minipage}{.475\hsize}\centering
\includegraphics[scale=.8]{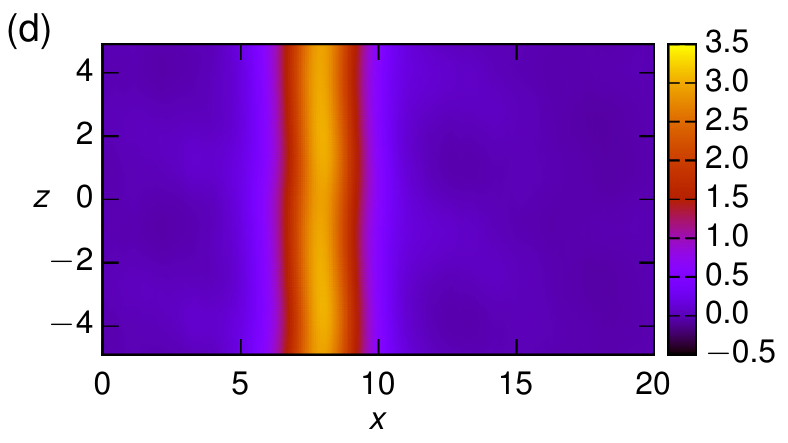}
\end{minipage}
\caption{$(y=0)$-cross-sections of $\phi$ with $a=0.06$ and $\kappa=2$ at different times.
(a) $t=25$,  (b) $t=32$, (c) $t=40$: the direction of deformation is changed with time.
(d) $t=60$: after the deformation, $\phi$ becomes homogeneous in $z$-direction, as in the initial state (Fig.~\ref{fig:line-soliton}b).}
\label{fig:phi_xz_a006}
\end{figure}

\begin{figure}[t]\centering
\includegraphics[scale=.9]{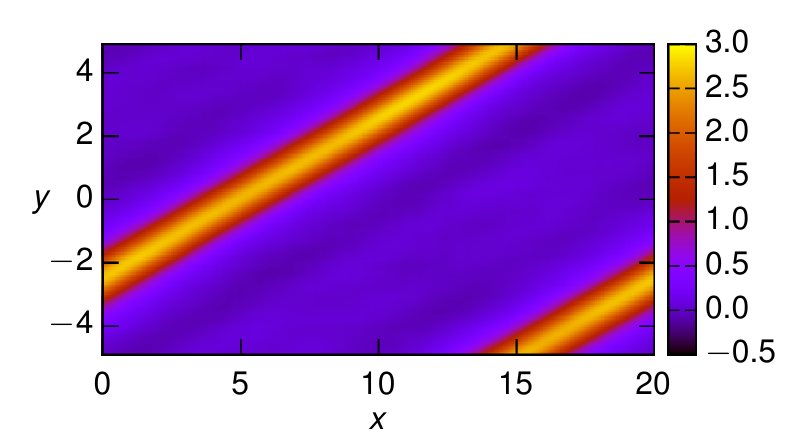}
\caption{$(z=0)$-cross-section of $\phi$ with $a=0.06$ and $\kappa=2$ at $t=32$.}{}
\label{fig:phi_xy_a006}
\end{figure}

%%%%%%%%%%%%%%%%%%%%%%%%%%%%%%%%%%%%%%%%%%%%%%%%%%%%%%%%%%%%%%%%
When $a=0$ (vortex-free), a line-soliton propagates without changing its shape.
Figures \ref{fig:phi_xz_a006} and \ref{fig:phi_xy_a006} show $y$- and $z$-cross-sections of $\phi$ with $a=0.06$ and $\kappa=2$.
In Fig.~\ref{fig:phi_xz_a006}, we observe deformations in the $y$-cross-sections:
(i) to the right side (Fig.~\ref{fig:phi_xz_a006}a); 
(ii) to both sides (Fig.~\ref{fig:phi_xz_a006}b); 
(iii) to the left side (Fig.~\ref{fig:phi_xz_a006}c);
(iv) after (i)--(iii), $\phi$ returns to the initial shape, homogeneous in the $z$-direction (Fig.~\ref{fig:phi_xz_a006}d).
We find that the deformations (i)--(iv) are repeated periodically.
In the $z$-cross-section, compared to the $y$-cross-section, noticeable deformations are not found (Fig.~\ref{fig:phi_xy_a006}).
From these results, we can say that line-solitons (homogeneous in the $z$-direction) are `stable' against weak vortexes.

%%%%%%%%%%%%%%%%%%%%%%%%%%%%%%%%%%%%%%%%%%%%%%%%%%%%%%%%%%%%%%%%
%%%%%%%%%%%%%%%%%%%%%%%%%%%%%%%%%%%%%%%%%%%%%%%%%%%%%%%%%%%%%%%%
%%%%%%%%%%%%%%%%%%%%%%%%%%%%%%%%%%%%%%%%%%%%%%%%%%%%%%%%%%%%%%%%
%%%%%%%%%%%%%%%%%%%%%%%%%%%%%%%%%%%%%%%%%%%%%%%%%%%%%%%%%%%%%%%%
%%%%%%%%%%%%%%%%%%%%%%%%%%%%%%%%%%%%%%%%%%%%%%%%%%%%%%%%%%%%%%%%
%%%%%%%%%%%%%%%%%%%%%%%%%%%%%%%%%%%%%%%%%%%%%%%%%%%%%%%%%%%%%%%%
%%%%%%%%%%%%%%%%%%%%%%%%%%%%%%%%%%%%%%%%%%%%%%%%%%%%%%%%%%%%%%%%

\begin{figure}[t]
\begin{minipage}{.475\hsize}\centering
\includegraphics[scale=.9]{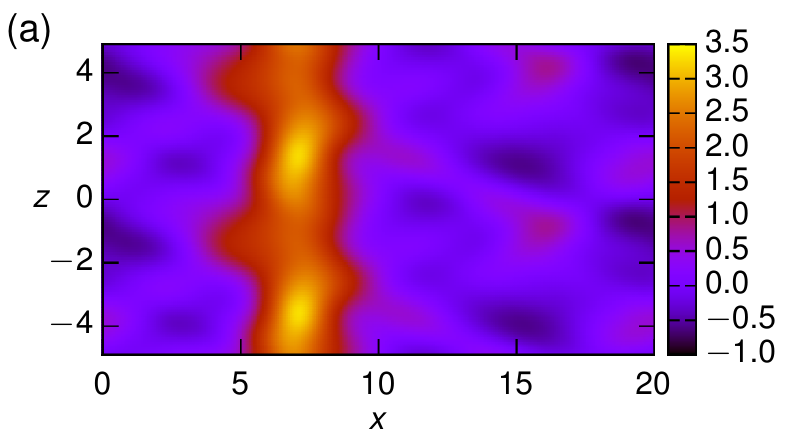}
\end{minipage}
\begin{minipage}{.475\hsize}\centering
\includegraphics[scale=.9]{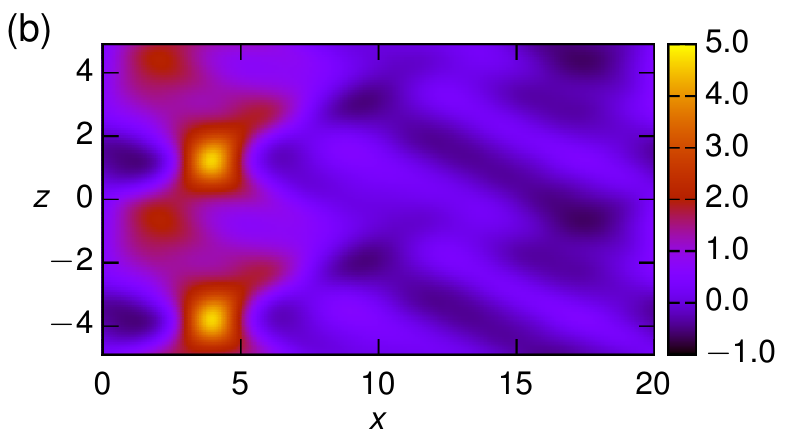}
\end{minipage}
\caption{$(y=0)$-cross-sections of $\phi$ with $a=0.30$ and $\kappa=2$ at different times: (a) $t=20$, (b) $t=40$.}
\label{fig:phi_xz_a030}
\end{figure}

\begin{figure}[t]\centering
\includegraphics[scale=.9]{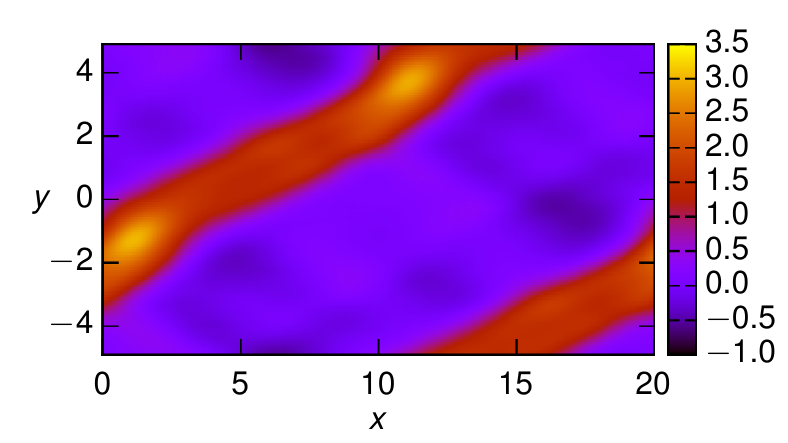}
\caption{$(z=0)$-cross-section of $\phi$ with $a=0.30$ and $\kappa=2$ at $t=40$.}
\label{fig:phi_xy_a030}
\end{figure}

\begin{figure}[t]
\begin{minipage}{.475\hsize}\centering
\includegraphics[scale=.9]{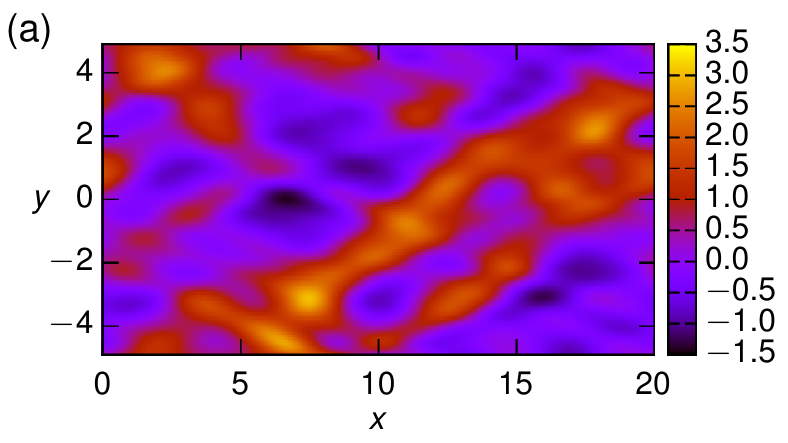}
\end{minipage}
\begin{minipage}{.475\hsize}\centering
\includegraphics[scale=.9]{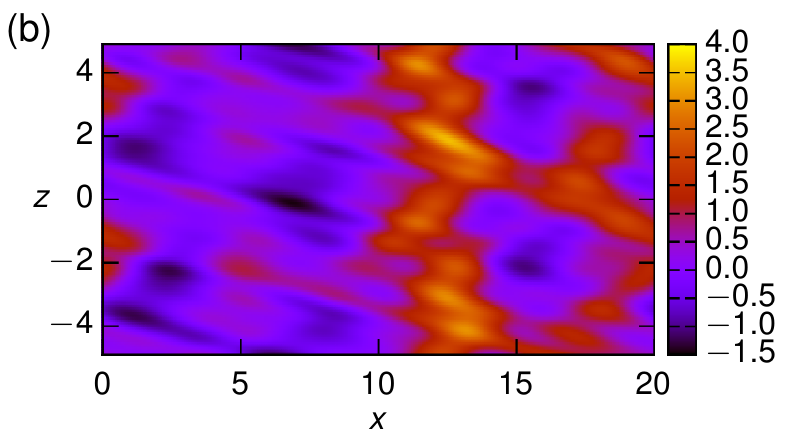}
\end{minipage}
\caption{Cross-sections of $\phi$ at $t=40$ with $a=1.0$ and $\kappa=2$: (a) ($z=0$)-cross-section, (b) ($y=0$)-cross-section.}
\label{fig:phi_a100}
\end{figure}

Figures \ref{fig:phi_xz_a030} and \ref{fig:phi_xy_a030} show $y$- and $z$-cross-sections of $\phi$ with a larger value of $a$ ($a=0.30$).
In the $y$-cross-section (Fig.~\ref{fig:phi_xz_a030}), We observe divided structures without returning to the initial shape.
Furthermore, differently from the result of $a=0.06$, deformations in the $z$-cross-section is also observed (Fig.~\ref{fig:phi_xy_a030}).
When $a$ is further large, $\phi$ breaks up into small structures and spreads, as found in Fig.~\ref{fig:phi_a100}.
These results show scatterings of line-solitons due to the ambient vortex fields, and they can be regarded as effects of the non-integrability.

%%%%%%%%%%%%%%%%%%%%%%%%%%%%%%%%%%%%%%%%%%%%%%%%%%%%%%%%%%%%%%%%
%%%%%%%%%%%%%%%%%%%%%%%%%%%%%%%%%%%%%%%%%%%%%%%%%%%%%%%%%%%%%%%%
%%%%%%%%%%%%%%%%%%%%%%%%%%%%%%%%%%%%%%%%%%%%%%%%%%%%%%%%%%%%%%%%
%%%%%%%%%%%%%%%%%%%%%%%%%%%%%%%%%%%%%%%%%%%%%%%%%%%%%%%%%%%%%%%%
%%%%%%%%%%%%%%%%%%%%%%%%%%%%%%%%%%%%%%%%%%%%%%%%%%%%%%%%%%%%%%%%
%%%%%%%%%%%%%%%%%%%%%%%%%%%%%%%%%%%%%%%%%%%%%%%%%%%%%%%%%%%%%%%%
%%%%%%%%%%%%%%%%%%%%%%%%%%%%%%%%%%%%%%%%%%%%%%%%%%%%%%%%%%%%%%%%
%%%%%%%%%%%%%%%%%%%%%%%%%%%%%%%%%%%%%%%%%%%%%%%%%%%%%%%%%%%%%%%%

\begin{figure}[t]\centering
\includegraphics[scale=.9]{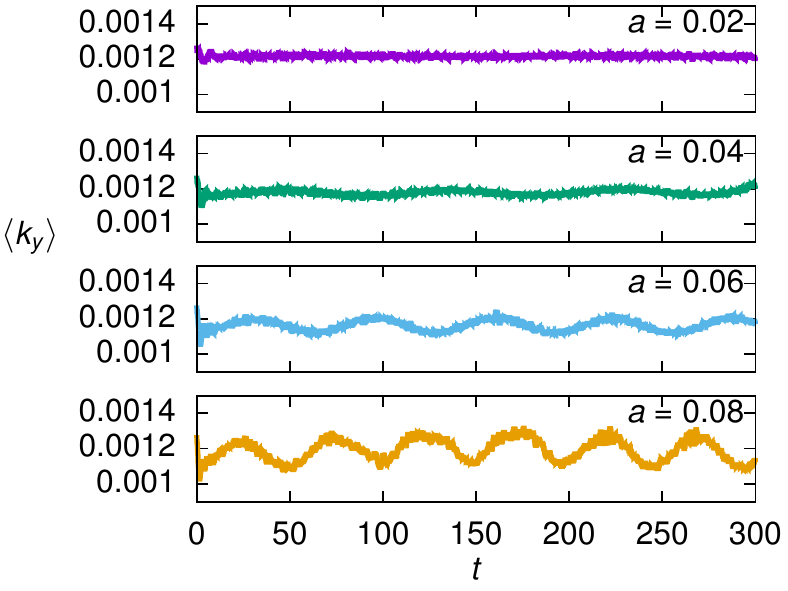}
\caption{Evolution of average wavenumber $\langle k_y \rangle$ with $a=0.02,0.04,0.06,0.08$. $\kappa=2$ is fixed.}
\label{fig:ky_small_a}
\end{figure}

\begin{figure}[t]\centering
\includegraphics[scale=.9]{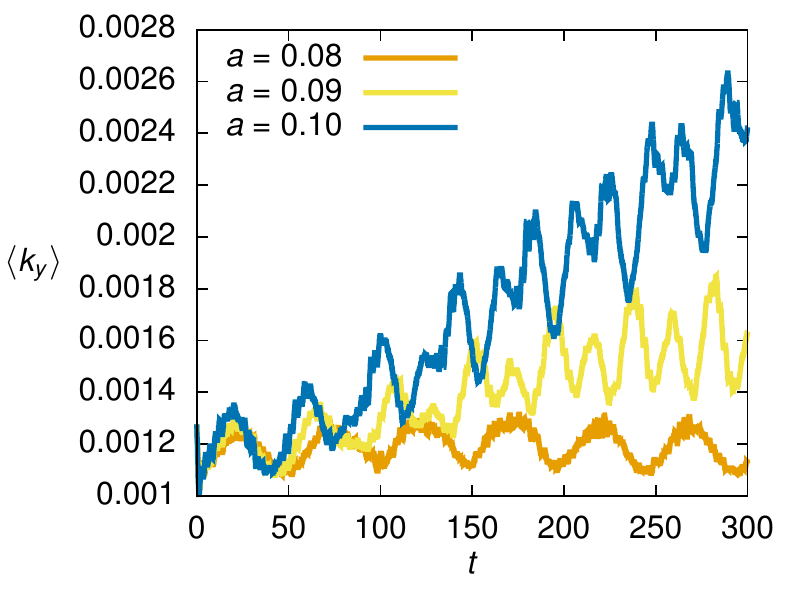}
\caption{Evolution of average wavenumber $\langle k_y \rangle$ with $a=0.08,0.09,0.10$. $\kappa=2$ is fixed.}
\label{fig:ky_middle_a}
\end{figure}

\begin{figure}[t]\centering
\includegraphics[scale=.9]{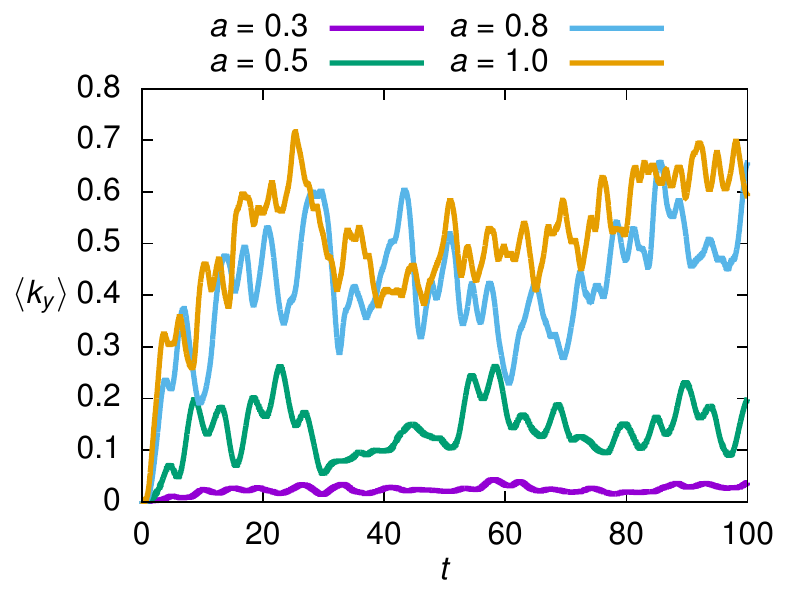}
\caption{Evolution of average wavenumber $\langle k_y \rangle$ with $a=0.3,0.5,0.8,1.0$. $\kappa=2$ is fixed.}
\label{fig:ky_large_a}
\end{figure}

\begin{figure}[t]\centering
\includegraphics[scale=.9]{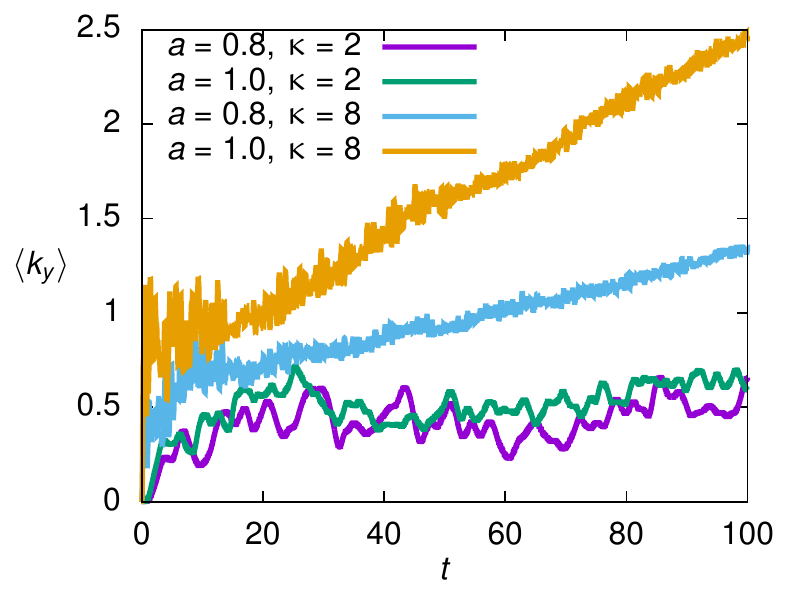}
\caption{Evolution of average wavenumber $\langle k_y \rangle$ with $(a,\kappa)=(0.8,2)$, $(1.0,2)$, %$(0.8,4)$, $(1.0,4)$, 
$(0.8,8)$, $(1.0,8)$.}
\label{fig:ky_large_ak}
\end{figure}

To evaluate the above observations quantitatively, we calculate the average wavenumber
\begin{equation}
    \langle k_j \rangle = \frac{\sum_{\bm k} k_j |\hat\phi_{\bm k}|^2}{\sum_{\bm k} |\hat\phi_{\bm k}|^2} \quad (j=x,y,z),
\end{equation}
where $\hat\phi_{\bm k}$'s are Fourier coefficients of $\phi$.
Figure \ref{fig:ky_small_a} shows the evolution of the average wavenumber $\langle k_y \rangle$ with $a=0.02$, $0.04$, $0.06$, and $0.08$ ($\kappa=2$ is fixed).
One can find that the evolution looks like periodic and the period becomes short when $a$ becomes large.
Figure \ref{fig:ky_middle_a} shows the evolution of $\langle k_y \rangle$ with $a=0.08$, $0.09$, and $0.10$.
A transition from the periodic behavior ($a=0.08$) to the increasing behavior ($a=0.09,0.10$) is found.
As shown in Fig.~\ref{fig:ky_large_a}, the average wavenumber has a large value when $a$ is further large.
In this stage, periodic evolutions are not observed.
One also finds that the value of the average wavenumber is not clearly different between in the case of $a=0.8$ and in that of $a=1.0$.
When $\kappa$ has a large value, the average wavenumber is found to grow larger (Fig.~\ref{fig:ky_large_ak}).
Thus, we can say that scattering scales of line-solitons depend on the intensity and the spatial scales of the ambient vortex fields.

%%%%%%%%%%%%%%%%%%%%%%%%%%%%%%%%%%%%%%%%%%%%%%%%%%%%%%%%%%%%%%%%
%%%%%%%%%%%%%%%%%%%%%%%%%%%%%%%%%%%%%%%%%%%%%%%%%%%%%%%%%%%%%%%%
%%%%%%%%%%%%%%%%%%%%%%%%%%%%%%%%%%%%%%%%%%%%%%%%%%%%%%%%%%%%%%%%
%%%%%%%%%%%%%%%%%%%%%%%%%%%%%%%%%%%%%%%%%%%%%%%%%%%%%%%%%%%%%%%%
%%%%%%%%%%%%%%%%%%%%%%%%%%%%%%%%%%%%%%%%%%%%%%%%%%%%%%%%%%%%%%%%
%%%%%%%%%%%%%%%%%%%%%%%%%%%%%%%%%%%%%%%%%%%%%%%%%%%%%%%%%%%%%%%%
%%%%%%%%%%%%%%%%%%%%%%%%%%%%%%%%%%%%%%%%%%%%%%%%%%%%%%%%%%%%%%%%
%%%%%%%%%%%%%%%%%%%%%%%%%%%%%%%%%%%%%%%%%%%%%%%%%%%%%%%%%%%%%%%%
%%%%%%%%%%%%%%%%%%%%%%%%%%%%%%%%%%%%%%%%%%%%%%%%%%%%%%%%%%%%%%%%

%%%%%%%%%%%%%%%%%%%%%%%%%%%%%%%%%%%%%%%%%%%%%%%%%%%%%%%%%%%%%%%%
%%%%%%%%%%%%%%%%%%%%%%%%%%%%%%%%%%%%%%%%%%%%%%%%%%%%%%%%%%%%%%%%
%%%%%%%%%%%%%%%%%%%%%%%%%%%%%%%%%%%%%%%%%%%%%%%%%%%%%%%%%%%%%%%%
\section{Conclusion}\label{sec:conclusion}
%%%%%%%%%%%%%%%%%%%%%%%%%%%%%%%%%%%%%%%%%%%%%%%%%%%%%%%%%%%%%%%%
%%%%%%%%%%%%%%%%%%%%%%%%%%%%%%%%%%%%%%%%%%%%%%%%%%%%%%%%%%%%%%%%
%%%%%%%%%%%%%%%%%%%%%%%%%%%%%%%%%%%%%%%%%%%%%%%%%%%%%%%%%%%%%%%%
The challenge of imparting vorticity to IAW was overcome by modifying the ordering of velocity field.
The newly formulated nonlinear system describes the scattering of IAWs propagating in the ambient vortex field.
The Painlev\'e test on the new system elucidates that the vorticity introduces essential three-dimensionality to the wave,
by which the integrability of the two-dimensional KP system is destroyed.
When the ambient vortex is weak, a two-dimensional line-soliton is deformed periodically but keep the solitary-wave structure.
This result indicates that a two-dimensional line-soliton is stable near the zero-vorticity state, even though the evolution equation is non-integrable.
The non-integrability (chaotic property) is found when the ambient vortex is strong;
line-solitons are broken into small scattered waves.

We end this paper with additional comments.
As we have shown in Section \ref{sec:KP_zeroV2},
the absence of vorticity is a strong imprint made by the ordering that characterizes the KP system.  
This constraint is ubiquitous among the families including a finite-temperature model (see \ref{sec:KP_baroc}),
trapped electron model, and multi-component models (see the references cited in Introduction).
% It is because the former only increases the number of equations and the latter only changes the Poisson equation, they do not modify the structure of the equation of motion.

The new ordering of the velocity field enables us to study the neighborhood of the integrable KP hierarchy.
At the lowest order, i.e., the KPY system, however, the range of dynamics is still rather narrow.
In fact, the \emph{helicity} $\int \bm u \cdot \bm\omega \, \rmd^3 x$
(the invariant characterizing the foliated phase space of general IAW \cite{YoshidaMorrison2016:phantom})
is zero for the KPY system; by equation \eqref{eq:constraint}, we find
\begin{equation}\label{eq:helicity}
\int \phi \lap_\perp \psi \, \rmd^3 x = \int \psi \left( \int \lap_\perp \phi \, \rmd x\right) \rmd y \, \rmd z = 0 .
\end{equation}
The generalized enstrophy $\int g(\Delta_\perp\psi) \, \rmd^2 x$ is also conserved ($g$ is an arbitrary function).
The constancy of this integral is not only due to the geometrical constraint (two-dimensionality) of $\psi$
but also because of the absence of the reciprocal reaction from the wave field $\phi$.
For a full development of turbulence, the enstrophy must be freed to increase, which is possible beyond the range of the present ordering.

% Thus, we can say that our model lives in a slightly generalized leaf around the singular leaf of the KP equation.
% The numerical results in Section \ref{sec:numerical} can be summarized with these words as followings.
% The line-solitons without ambient vortexes lives in the singular leaf. 
% Dynamical orbits starting from the points close to the leaf (with weak vortexes) stay around moving periodically, and line-solitons keep their ordered solitary structures.
% Those starting from the points far from the leaf (with strong vortexes) leave far away, and line-solitons scatter into small structures.

%%%%%%%%%%%%%%%%%%%%%%%%%%%%%%%%%%%%%%%%%%%%%%%%%%%%%%%%%%%%%%%%
%%%%%%%%%%%%%%%%%%%%%%%%%%%%%%%%%%%%%%%%%%%%%%%%%%%%%%%%%%%%%%%%
%%%%%%%%%%%%%%%%%%%%%%%%%%%%%%%%%%%%%%%%%%%%%%%%%%%%%%%%%%%%%%%%

\section*{Acknowledgments}
The authors acknowledge stimulating discussions with Dr.\ Hosam Abd El Razek during his visit to The University of Tokyo.
This research was supported by JSPS KAKENHI Grant Numbers 23224014 and 15K13532.

%%%%%%%%%%%%%%%%%%%%%%%%%%%%%%%%%%%%%%%%%%%%%%%%%%%%%%%%%%%%%%%%
%%%%%%%%%%%%%%%%%%%%%%%%%%%%%%%%%%%%%%%%%%%%%%%%%%%%%%%%%%%%%%%%
%%%%%%%%%%%%%%%%%%%%%%%%%%%%%%%%%%%%%%%%%%%%%%%%%%%%%%%%%%%%%%%%
\appendix
\section{Finite ion temperature effect}\label{sec:KP_baroc}
%%%%%%%%%%%%%%%%%%%%%%%%%%%%%%%%%%%%%%%%%%%%%%%%%%%%%%%%%%%%%%%%
In the vorticity equation \eqref{eq:vort_boost}, the boost term $\pd\bm\omega/\pd x$ causes unbalance of ordering
and forces the vorticity to vanish, when we invoke the standard expansion \eqref{eq:rp}
(this ordering is tailor-made to match $n(\nabla \cdot \bm v)$ and $\nabla\phi$ with the boost terms 
in the continuity equation and the equation of motion).

In this appendix, we examine the effect of finite ion temperature $T_\rmi$.
This introduces a non-potential force $- n^{-1} \nabla p$ to the equation of motion \eqref{eq:mom} 
and a source term $\nabla T_\rmi \times \nabla s$ to the vorticity equation \eqref{eq:vort}, where $p$ is the ion pressure and $s$ is the ion entropy.
This production mechanism of vorticity is called the \emph{baroclinic effect} \cite{Pedlosky,DelSordo2011}.
We show that the baroclinic effect must be absent for the success of the reductive perturbation method.

We apply the reductive perturbation method for the KP equation (Section \ref{sec:KP_deriv}).
We assume that the ion pressure is governed by the adiabatic equation
\begin{equation}\label{eq:press}
    \pdif{p}{t} + \bm u \cdot \nabla p + \Gamma p (\nabla \cdot \bm u) = 0,
\end{equation}
where $\Gamma$ is the heat ratio.
The pressure is normalized by the representative pressure $n_0 T_\rme$.
We expand $p$ as $p = \sigma + \epsilon^2 p_1 + \epsilon^4 p_2 + \cdots{}$, 
where $\sigma = T_{\rmi0}/T_\rme$ is the normalized representative temperature.
It should be noted that the speed of Galilean boost must be modified from $1$ to $\lambda = \sqrt{1+\Gamma\sigma}$ 
(normalized with $c_\rms = \sqrt{T_\rme/m_\rmi}$) \cite{Tagare1973:KdV}.
The relations \eqref{eq:KP.e1}, \eqref{eq:KP.e2y}, and \eqref{eq:KP.e2z} are modified as
\begin{equation}\label{eq:tKP.e12}
\left\{ \begin{aligned}
    \lambda n_1 &= \lambda \phi_1 = u_1 = \frac{\phi_1+p_1}{\lambda}, \\
    \pdif{v_1}{\tilde x} &= \frac{1}{\lambda} \pdif{}{\tilde y} (\phi_1+p_1), \\
    \pdif{w_1}{\tilde x} &= \frac{1}{\lambda} \pdif{}{\tilde z} (\phi_1+p_1),
\end{aligned} \right.
\end{equation}
and we obtain the KP equation
\begin{equation}
    \pdif{}{\tilde x} \left[ \pdif{\phi_1}{\tilde t} + \left(1 + \frac{\Gamma+1}{2}\Gamma\sigma \right) \phi_1 \pdif{\phi_1}{\tilde x} + \frac{1}{2\lambda} \pdif[3]{\phi_1}{\tilde x} \right] + \frac{\lambda}{2} \lap_\perp \phi_1 = 0,
\end{equation}
which reduces to equation \eqref{eq:KP_3D} in the limit $T_\rmi = 0$ ($\sigma=0$, $\lambda = 1$).

From equation \eqref{eq:tKP.e12}, we find that the lowest order vorticity is zero: $\omega_{x1}=\omega_{y1}=\omega_{z1} = 0$, same as the case of cold ions ($T_\rmi=0$).
Furthermore, we show that entropy must be homogeneous and thus baroclinic term must vanishes for the success of the reductive perturbation method.
We use the same procedure of Section \ref{sec:KP_zeroV2}.
Let us consider the adiabatic evolution equation for entropy
\begin{equation}\label{eq:entropy}
    \pdif{s}{t} + \bm u \cdot \nabla s = 0,
\end{equation}
which is equivalent to the pressure equation \eqref{eq:press}.
The Galilean boost in the $x$-direction modifies equation \eqref{eq:entropy} as
\begin{equation}\label{eq:entropy_boost}
    \pdif{s}{t} - \lambda \pdif{s}{x} + \bm u \cdot \nabla s = 0.
\end{equation}
Now we evaluate the orders of operators $\pd/\pd t$, $\lambda \pd/\pd x$, and $\bm u \cdot \nabla = u \pd/\pd x + v \pd/\pd y + w \pd/\pd z$ with equations \eqref{eq:rp_xyt}, \eqref{eq:rp}, and \eqref{eq:rp_z}.
The order of the second operator is $\epsilon^1$, and the lowest order of others is $\epsilon^3$.
Thus, the leading order of entropy $s_0$ must satisfy $\pd s_0/\pd x = 0$.
This results in $s_0 = c$ (constant) under the boundary condition $s_0 \to c$ $(x \to \pm \infty)$, 
which is a natural choice because $s_0$ is not a perturbation part.
Eliminating $s_0$, equation \eqref{eq:entropy_boost} reads as the equation for $s_1$.
%Now the minimum order of each operator increases by $\epsilon^2$.
%Then the same discussion is valid for $s_1$, and then $s_1$ must satisfy $\pd s_1/\pd x = 0$.
The same discussion requires $s_1$ to satisfy $\pd s_1/\pd x = 0$.
Since $s_1$ is a perturbation part, it is natural to use the boundary condition $s_1 \to 0$ $(x \to \infty)$.
These relations lead to $s_1 = 0$.
Repeating this procedure, we find that entropy must be homogeneous: $s = s_0 = c$.
%Thus, the baroclinic effect must vanish under the reductive perturbation method.

We note that baroclinic effect vanishes even in our finite-vorticity system (Section \ref{sec:KPY}).
This is because the above discussion is valid even if we introduce the additional velocity $\bm v_0$ as equation \eqref{eq:rp_KPY}.
Finite ion temperature modifies the KPY equation \eqref{eq:KPY} as
\begin{equation}
    \pdif{}{\tilde x} \left[
        \pdif{\phi_1}{\tilde t} + \left(1 + \frac{\Gamma+1}{2}\Gamma\sigma \right) \phi_1 \pdif{\phi_1}{\tilde x} 
        + \frac{1}{2\lambda} \pdif[3]{\phi_1}{\tilde x} + [\phi_1, \psi] \right] 
    + \frac{\lambda}{2} \lap_\perp \phi_1 = 0,
\end{equation}
without changing the Euler vorticity equation \eqref{eq:2DEV}.

%%%%%%%%%%%%%%%%%%%%%%%%%%%%%%%%%%%%%%%%%%%%%%%%%%%%%%%%%%%%%%%%
%%%%%%%%%%%%%%%%%%%%%%%%%%%%%%%%%%%%%%%%%%%%%%%%%%%%%%%%%%%%%%%%
%%%%%%%%%%%%%%%%%%%%%%%%%%%%%%%%%%%%%%%%%%%%%%%%%%%%%%%%%%%%%%%%
\bibliography{kpy_paper}

\end{document}